\documentclass[twoside,slac_one]{revtex4}
\usepackage{graphicx}
\usepackage{fancyhdr}
\usepackage{amsmath} 
\usepackage{bm}
\usepackage{amsxtra}
\usepackage{amssymb}
\usepackage{amsthm}
\usepackage{latexsym}
\usepackage{lscape}

\def\lsim{\mathrel{\rlap{\lower4pt\hbox{\hskip1pt$\sim$}}
    \raise1pt\hbox{$<$}}}
\def\gsim{\mathrel{\rlap{\lower4pt\hbox{\hskip1pt$\sim$}}
    \raise1pt\hbox{$>$}}} 
    
\newcommand{\slashed}[1]{\setbox0=\hbox{$#1$}   
   \dimen0=\wd0                                     
   \setbox1=\hbox{/} \dimen1=\wd1                   
   \ifdim\dimen0>\dimen1                            
      \rlap{\hbox to \dimen0{\hfil/\hfil}}          
      #1                                            
   \else                                            
      \rlap{\hbox to \dimen1{\hfil$#1$\hfil}}       
      /                                             
   \fi}                                             %

\begin{document}

\title{SQCD Corrections to $bg \rightarrow bh$}

%

\author{S.~Dawson$^{a}$, C.~B.~Jackson$^{b}$, P.~Jaiswal$^{a,c}$}
\affiliation{
$^a$Department of Physics, Brookhaven National Laboratory, 
Upton, NY 11973, USA \\
$^{b}$Physics Department, University of Texas, Arlington, Texas \\
$^{c}$Yang Institute for Theoretical Physics, Stony Brook University, Stony Brook, NY 11790, USA
}

\begin{abstract}
In the Minimal Supersymmetric Standard Model, the effective $b$ quark Yukawa coupling to the lightest neutral Higgs boson is enhanced. Therefore, the associated production of the lightest Higgs boson with a $b$ quark is an important discovery channel. We consider the SUSY QCD contributions from squarks and gluinos and discuss the decoupling properties of these effects. A comparision of our exact ${\cal O}(\alpha_s)$ results with those of a widely used effective Lagrangian approach, the $\Delta_b$ approximation, is also presented.  
\end{abstract}

\maketitle

\thispagestyle{fancy}


\section{Introduction}

In the MSSM, the
production mechanisms for the Higgs bosons can be significantly different from
 that in the Standard Model. For large values of $\tan\beta$, the heavier Higgs
bosons, $A$ and $H$,
are predominantly produced in association with $b$ quarks.  
Even for 
$\tan\beta\sim 5$, the production rate in association with $b$ quarks is
similar to that from gluon fusion for $A$ and $H$
production (\cite{Dittmaier:2011ti}). For the lighter Higgs
boson, $h$, for
$\tan\beta \gsim 7$ the dominant production mechanism 
 at both the Tevatron and the LHC
 is production with $b$ quarks for light $M_A$ ($\lsim 200~GeV$), where 
the $b {\overline b} h$ coupling is enhanced .
Both the Tevatron (\cite{Benjamin:2010xb})
and the LHC experiments (\cite{Chatrchyan:2011nx})
have presented limits Higgs production in association
with $b$ quarks, searching for the 
decays $h\rightarrow \tau^+\tau^-$ and $b \overline{b}$. 
  These limits 
are obtained in the context of the MSSM are sensitive 
to the $b$-squark and gluino
loop corrections which we consider here.

The rates for $bh$ associated production at
the LHC and the Tevatron 
have been extensively studied (\cite{Dawson:2005vi, Campbell:2004pu, Maltoni:2003pn, Dawson:2004sh, Dittmaier:2003ej, Dicus:1998hs, Dawson:2003kb, Maltoni:2005wd, Campbell:2002zm, Carena:2007aq, Carena:1998gk}) and the NLO QCD correction
are well understood, 
both in the $4$- and $5$- flavor number parton 
schemes (\cite{Dawson:2005vi, Campbell:2004pu, Maltoni:2003pn}).  
In the $4$- flavor number scheme, the lowest order processes 
for producing a Higgs boson and a $b$ quark are $gg\rightarrow
 b {\overline b}h$ and 
$ q {\overline q}\rightarrow b {\overline b} 
h$ (\cite{Dittmaier:2003ej, Dawson:2004sh, Dicus:1998hs}).  In the $5-$ flavor
number scheme, the lowest order process is $b g \rightarrow b h$
 (${\overline {b}} g \rightarrow {\overline {b}}h$).
The two schemes represent different orderings of perturbation theory and
  calculations
in the two schemes produce rates which are in qualitative 
agreement (\cite{Campbell:2004pu, Dittmaier:2011ti}).  
In this paper, we use the $5$-flavor number scheme for simplicity. 
The resummation of threshold logarithms (\cite{Field:2007ye}), 
electroweak corrections (\cite{Dawson:2010yz,Beccaria:2010fg})
 and SUSY QCD corrections (\cite{Dawson:2007ur}) have also been
computed for $bh$ production in the $5-$ flavor number scheme.
  
Here, we focus on the role of squark and gluino loops.  
The properties of the SUSY QCD corrections to the $b {\overline b}
 h$ vertex, both
for the decay $h\rightarrow b {\overline b}
$ (\cite{Dabelstein:1995js, Hall:1993gn, Carena:1999py, Guasch:2003cv})
and the production, 
$b {\overline b}\rightarrow h$ (\cite{Haber:2000kq, Harlander:2003ai, Guasch:2003cv, Dittmaier:2003ej}), were computed
long ago.  
The contributions from $b$ squarks and gluinos to the
lightest MSSM Higgs boson
mass are known at $2$-loops (\cite{Heinemeyer:2004xw,Brignole:2002bz}), 
while the $2$-loop
SQCD contributions to the $b{\overline b}h$ vertex
 is known in the limit in which the
Higgs mass is much smaller than the squark and gluino 
masses (\cite{Noth:2010jy,Noth:2008tw}).
The contributions of squarks and gluinos to the 
on-shell $b {\overline b}
h$ vertex are non-decoupling for heavy squark and gluino masses and
decoupling is only achieved when the pseudoscalar
mass, $M_A$, also becomes large.

An effective Lagrangian approach, the $\Delta_b$ 
approximation (\cite{Carena:1999py, Hall:1993gn}),
can be used to approximate 
the SQCD contributions to 
the on-shell $b {\overline b}h$ vertex
and to resum the $(\alpha_s  \tan\beta/M_{SUSY})^n$ enhanced terms.
The numerical accuracy of the $\Delta_b$ effective Lagrangian approach
has been examined for a number of cases.  
The $2-$loop contributions to
the lightest MSSM Higgs boson mass 
of ${\cal O}(\alpha_b\alpha_s)$ were computed by 
\cite{Heinemeyer:2004xw} and \cite{Brignole:2002bz}, 
and it was found that the majority of these corrections
could be absorbed into a $1-$loop 
contribution by defining an effective  $b$ quark
 mass using the $\Delta_b$ approach. 
The sub-leading contributions to the Higgs
boson mass (those not absorbed into $\Delta_b$)
are then of ${\cal O}(1~GeV)$.  
The $\Delta_b$ approach  also yields an excellent
approximation to the SQCD corrections for
 the decay process $h\rightarrow b
{\overline b}$ (\cite{Guasch:2003cv}).  
It is particularly interesting
to study the 
accuracy of the
$\Delta_b$ approximation 
for production processes where one of the $b$ quarks is off-shell.
The SQCD contributions from squarks and gluinos to the 
inclusive Higgs production rate in association with $b$ quarks has been
studied extensively in the 4FNS by \cite{Dittmaier:2006cz}, where the the lowest order
contribution is $gg\rightarrow b {\overline b}h$.  
In the 4FNS, the inclusive
cross section including the exact 1-loop SQCD corrections is reproduced
to within a few percent using the $\Delta_b$ approximation.  However,
the accuracy of the $\Delta_b$ approximation for the MSSM
neutral Higgs boson 
production in the
5FNS has been studied for only a small set of MSSM parameters in 
Ref. \cite{Dawson:2007ur}.
The major new result of this paper is a detailed study of
the accuracy of the $\Delta_b$ approach in the 5FNS 
for the $bg\rightarrow bh$ production process.  In this case,
one of the $b$ quarks is off-shell and there are contributions which are not
contained in the effective Lagrangian approach. 

In this article, we give a brief review of 
the effective Lagrangian approximation in section 1.  
In section 2, we summarize the SQCD calculations for $bg \rightarrow bh$ (\cite{Dawson:2007ur}) including  terms which are enhanced by
 $m_b \tan\beta$ (\cite{Dawson:2011pe}). 
Analytic results for the SQCD corrections
to $bg\rightarrow bh$ in the extreme mixing scenarios in the $b$ squark
sector have been calculated by \cite{Dawson:2011pe} and are presented 
 in Section 3.  
Section 4 contains
numerical results for the $\sqrt{s}=7$ TeV
LHC.  Finally, our conclusions are summarized in Section 5.

\section{SQCD Contributions to $g b\rightarrow b h$}

\subsection{$\Delta_{b}$
 Approximation:  The Effective Lagrangian Approach} 
\label{sec:db}
Loop corrections which are enhanced by powers of $\alpha_s\tan\beta$
can be included in an
effective Lagrangian approach (\cite{Carena:1999py, Hall:1993gn, Guasch:2003cv}).
Using the effective Lagrangian, 
which we term the Improved Born
Approximation (or
$\Delta_b$ approximation), 
the cross section is written in terms of the effective
coupling,
\begin{equation}
g_{bbh}^{\Delta_b}\equiv g_{bbh}\biggl({1\over 1+\Delta_b}\biggr)
\biggl(1-{\Delta_b\over \tan\beta
\tan \alpha}\biggr) \, ,
\label{effcoup}
\end{equation}
where
\begin{equation}
g_{bbh}=-\biggl({\sin\alpha\over \cos\beta}\biggr)
{{\overline m_b}(\mu_R)\over v_{SM}}
\end{equation}
and  the
$1$-loop
contribution to $\Delta_b$ from sbottom/gluino loops 
is (\cite{Carena:1994bv,Hall:1993gn,Carena:1999py}) 
\begin{equation}
\Delta_b={2\alpha_s(\mu_S)\over 3 \pi} M_{\tilde g} \mu 
\tan\beta
I(M_{\tilde {b_1}},
M_{\tilde{ b_2}}, M_{\tilde g})\, ,
\label{db}
\end{equation}
where the function $I(a,b,c)$ is,
\begin{equation}
I(a,b,c)={1\over (a^2-b^2)(b^2-c^2)(a^2-c^2)}\biggl\{a^2b^2\log\biggl({a^2\over b^2}\biggr)
+b^2c^2\log\biggl({b^2\over c^2}\biggr)
+c^2a^2\log\biggl({c^2\over a^2}\biggr)\biggr\}\, ,
\end{equation}
The Improved Born Approximation consists of rescaling the tree level
cross section, $\sigma_0$, by the coupling of 
Eq. \ref{effcoup},
\begin{equation}
\sigma_{IBA}=\biggl({g_{bbh}^{\Delta_b}\over g_{bbh}}\biggr)^2\sigma_0
\, .
\label{sigibadef}
\end{equation}
The Improved Born Approximation has been shown
to accurately reproduce the full
SQCD calculation of
$pp \rightarrow {\overline t} b H^+$ (\cite{Berger:2003sm,Dittmaier:2009np}).
The one-loop result including the SQCD corrections
for $b g\rightarrow b h$  can be written as,
\begin{eqnarray}
\sigma_{SQCD}&\equiv&
\sigma_{IBA}\biggl(1+\Delta_{SQCD}\biggr)
\, ,
\end{eqnarray}
where $\Delta_{SQCD}$ is found from
the exact SQCD
calculation summarized in Appendix B of (\cite{Dawson:2011pe}).

\subsection{Full One-loop SQCD Contributions to $g b\rightarrow b h$}

The SQCD contributions to the $g b\rightarrow b h$ process 
have been computed in Ref. (\cite{Dawson:2007ur}) in the $m_b=0$ limit and further, 
in Ref. (\cite{Dawson:2011pe}) where terms which are enhanced
by $m_b \tan\beta$ have been included.  

\begin{figure}[t]
\begin{center}
\includegraphics[scale=0.5]{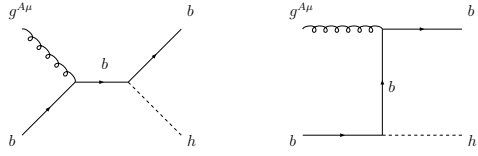} 
\caption[]{Feynman diagrams for $ g(q_1)+b (q_2)\rightarrow
b(p_b)+ h(p_h)$.}
\label{fg:bghb_feyn}
\end{center}
\end{figure}

The tree level diagrams for $g(q_1) + b(q_2) \to b(p_b) + h(p_h)$
are shown in Fig. \ref{fg:bghb_feyn}.
The amplitude can be written as a sum of following dimensionless spinor products 
\begin{eqnarray}
M_{s}^{\mu} & = & \frac{\overline{u}\left(p_{b}\right)\left(\slashed{q}_{1}+\slashed{q}_{2}\right)\gamma^{\mu}u\left(q_{2}\right)}{s}\nonumber \\
M_{t}^{\mu} & = & \frac{\overline{u}\left(p_{b}\right)\gamma^{\mu}\left(\slashed{p}_{b}-\slashed{q}_{1}\right)u\left(q_{2}\right)}{t}\nonumber \\
M_{1}^{\mu} & = & q_{2}^{\mu}\frac{\overline{u}\left(p_{b}\right)u\left(q_{2}\right)}{u}\nonumber \\
M_{2}^{\mu} & = & \frac{\overline{u}\left(p_{b}\right)\gamma^{\mu}u\left(q_{2}\right)}{m_{b}}\nonumber \\
M_{3}^{\mu} & = & p_{b}^{\mu}\frac{\overline{u}\left(p_{b}\right)\slashed{q}_{1}u\left(q_{2}\right)}{m_{b}t}\nonumber \\
M_{4}^{\mu} & = & q_{2}^{\mu}\frac{\overline{u}\left(p_{b}\right)\slashed{q}_{1}u\left(q_{2}\right)}{m_{b}s}\, ,
\label{eq: SME}
\end{eqnarray}
where $s=(q_1+q_2)^2, t=(p_b-q_1)^2$ and $u=(p_b-q_2)^2$.
In the  $m_b=0$ limit, the tree level amplitude depends 
only on $M_s^\mu$ and $M_t^\mu$, and $M_1^\mu$ is
generated at one-loop. 
When the effects of the $b$ mass are included, $M_2^\mu$, $M_3^\mu$, and $M_4^\mu$ 
are also generated.

The tree level amplitude is 
\begin{eqnarray}
\mathcal{A}_{\alpha\beta}^{a}\mid_0 & = & -g_{s}
g_{bbh}\left(T^{a}\right)_{\alpha\beta}\epsilon_{\mu}(q_1)\left\{ M_{s}^{\mu}+M_{t}^{\mu}\right\} \, ,
\end{eqnarray}
and the one loop contribution can be written as 
\begin{equation}
\mathcal{A}_{\alpha\beta}^{a}
=-\frac{\alpha_{s}(\mu_R)}{4\pi}g_{s}g_{bbh}\left(T^{a}\right)_{\alpha\beta}\sum_{j}X_{j}M_{j}^{\mu}\epsilon_{\mu}(q_1)\, .
\label{onedef}
\end{equation}
For detailed calculation of counter-terms and the coefficients $X_j$, cf. (\cite{Dawson:2011pe}).

\section{Results for Maximal and Minimal Mixing in the $b$-Squark Sector}

\subsection{Maximal Mixing}

The SQCD contributions to $bg\rightarrow bh$
can be examined analytically in several
scenarios.  In the maximal mixing scenario, 
\begin{equation}
\mid {\tilde m}_L^2
-{\tilde m}_R^2\mid << {m_b\over 1+\Delta_b}
\mid X_b\mid\, .
\end{equation} 
 We expand in powers
of   ${\mid {\tilde m}_L^2
-{\tilde m}_R^2\mid \over m_b X_b}$.  In this case the sbottom masses
are nearly degenerate,
\begin{eqnarray}
 M_S^2&\equiv &{1\over 2}
\biggl[
M_{{\tilde b}_1}^2+M_{{\tilde b}_2}^2
\biggr]
\nonumber \\
\mid M_{{\tilde b}_1}^2-M_{{\tilde b}_2}^2\mid
&=&
\biggl({2m_b \mid X_b\mid \over 1+\Delta_b}\biggr)
\biggl(1+{( {\tilde m}_L^2
-{\tilde m}_R^2)^2 (1+\Delta_b)^2\over 8  m_b^2 X_b^2}\biggr)<< M_S^2
\, .
\end{eqnarray}
This scenario is termed maximal mixing since
\begin{equation}
\sin 2{{\tilde{\theta}}}_b\sim 1-
{({\tilde m}_L^2
-{\tilde m}_R^2)^2(1+\Delta_b)^2\over 8  m_b^2 X_b^2}
\, .
\end{equation}
We expand the contributions of
the exact one-loop SQCD calculation (see Appendix B of \cite{Dawson:2011pe}) in powers of $1/M_S$, keeping terms
to ${\cal O}\biggl({M_{EW}^2\over M_S^2}\biggr)$ and assuming
$M_S\sim M_{\tilde g}\sim \mu\sim A_b\sim
{\tilde m}_L\sim {\tilde m}_R >> M_W, M_Z, M_h\sim M_{EW}$.
In the expansions, we assume the
large $\tan\beta$ limit and take 
$m_b\tan\beta\sim {\cal {O}}(M_{EW})$.
This expansion has been studied in
detail for the decay 
$h\rightarrow b {\overline{b}}$, 
with particular emphasis
on the decoupling properties of the results as 
$M_S$ and $M_{\tilde g}
\rightarrow\infty$ (\cite{Haber:2000kq}).
Recently, we studied this expansion of SQCD contributions to the process, $bg\rightarrow bh$ in Ref (\cite{Dawson:2011pe}). Here, we present the final results of the calculation. The amplitude squared, summing over final state spins and colors and averaging
 over initial state spins and colors, including one-loop SQCD
 corrections is 
\begin{eqnarray}
\left|
{\overline{\mathcal{A}}}\right|_{max}^{2} 
 & = & -\frac{2\pi\alpha_{s}(\mu_R)}{3}(g_{bbh}^{\Delta_b})^{2}
\left\{\left(
\frac{u^{2}+M_{h}^{4}}{st}\right)\left[1+2
\biggl({\delta g_{bbh}\over g_{bbh}}\biggr)^{(2)}_{max}
\right]\right.
\nonumber \\ &&
\left.+{\alpha_s(\mu_R)\over 2\pi}
\frac{M_{h}^{2}}{M_{S}^{2}}\delta\kappa_{max}\right\}
+{\cal O}\biggl(\biggl[{M_{EW}\over M_S}\biggr]^4,\alpha_s^3\biggr)\, .
\label{maxans}
\end{eqnarray}
where,  $g_{bbh}^{\Delta_b}$ is the rescaled bottom quark Yukawa coupling and the subleading terms, $\biggl({\delta g_{bbh}\over g_{bbh}}\biggr)^{(2)}_{max}$ and $\delta\kappa_{max} \frac{M_h^2}{M_S^2}$ of ${\cal O}(M_{EW}^2/M_S^2)$ are given in (\cite{Dawson:2011pe}). These subleading terms are not included in the IBA approximation and their conribution is usually small.

\subsection{Minimal Mixing}
The minimal mixing scenario is characterized by a mass splitting between the $b$ squarks
which is of order the $b$ squark mass,
$\mid M_{\tilde{b}_1}^2-M_{\tilde{b}_2}^2\mid \sim M_S^2$.  
In this case, 
\begin{equation}
\mid {\tilde m}_L^2-{\tilde m}_R^2\mid >> {m_b
\mid X_b\mid\over (1+\Delta_b)}\, ,
\end{equation} 
and the mixing angle in the $b$ squark sector is close to zero,
\begin{equation}
\cos 2 {\tilde \theta}_b\sim 1-{2m_b^2X_b^2\over 
(M_{{\tilde b}_1}^2-M_{{\tilde b}_2}^2)^2}
\biggl({1\over 1+\Delta_b}\biggr)^2
\, .
\end{equation}

As in the previous section, the  spin and color 
averaged amplitude-squared is,
\begin{eqnarray}
\mid {\overline A}\mid_{min}^2 
&=&-{2\alpha_s\pi\over 3}(g_{bbh}^{\Delta_b})^2\biggl\{
{(M_h^4+u^2)\over s t}\biggl[1+2\biggl({\delta g_{bbh}\over
g_{bbh}}\biggr)^{(2)}_{min}\biggr]
\nonumber \\ && 
+{\alpha_s\over 2\pi}\delta \kappa_{min}
{M_h^2\over M_{\tilde g}^2}
\biggr\}+{\cal O}\biggl(\biggl[{M_{EW}\over M_S}\biggr]^4,\alpha_s^3\biggr)
\, .
\label{minansdef}
\end{eqnarray}
The contributions which are not contained in $\sigma_{IBA}$, $\biggl({\delta g_{bbh}\over g_{bbh}}\biggr)^{(2)}_{min}$ and $\delta\kappa_{min} \frac{M_h^2}{M_S^2}$ are given in (\cite{Dawson:2011pe}) and
again found to be
suppressed by ${\cal O}\biggl(\biggl[{M_{EW}\over M_S}\biggr]^2\biggr)$.

\section{Numerical Results}

\begin{figure}[h]
\begin{center}
\vspace{2mm}
\includegraphics[scale=0.45]{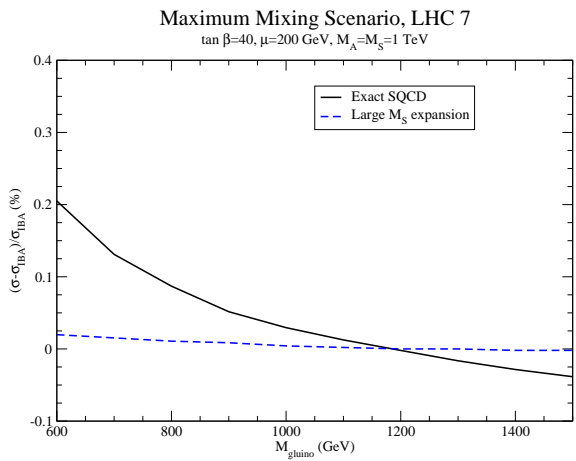} 
\caption[]{Percentage difference between the Improved Born Approximation and
the exact one-loop SQCD calculation of $pp\rightarrow
bh$ for maximal mixing in the $b$-squark 
sector at $\sqrt{s}=7~TeV$, $\tan\beta=40$, and $M_A=1~TeV$.}
\label{fg:maxmix}
\end{center}
\end{figure}

\begin{figure}[h]
\begin{center}
\vspace{2mm}
\includegraphics[scale=0.45]{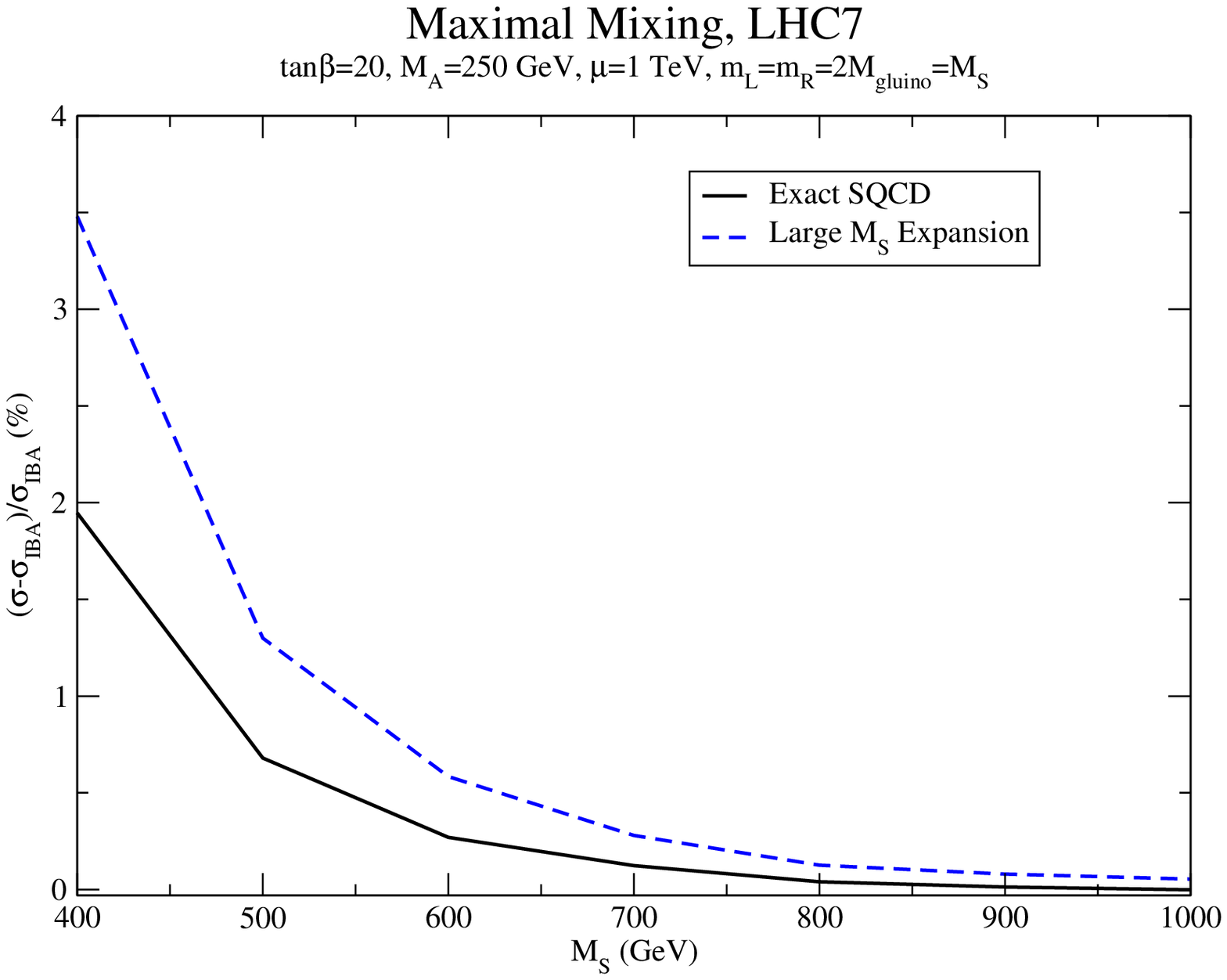} 
\caption[]{Percentage difference between the Improved Born Approximation and
the exact one-loop SQCD calculation of $pp\rightarrow
bh$ for maximal mixing in the $b$-squark 
sector at $\sqrt{s}=7~TeV$, $\tan\beta=20$, and $M_A=250~GeV$.}
\label{fg:maxmix250}
\end{center}
\end{figure}

\begin{figure}[h]
\begin{center}
\vspace{2mm}
\includegraphics[scale=0.45]{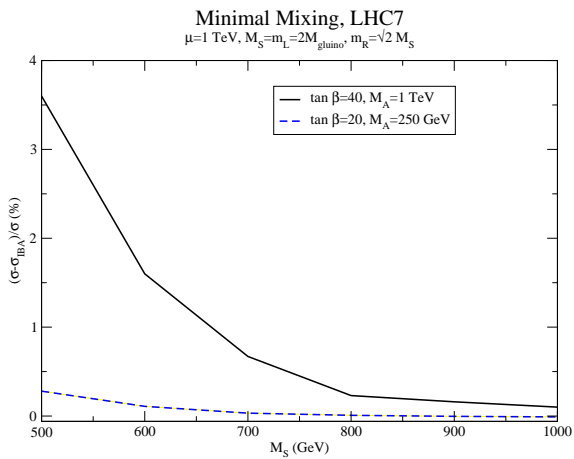} 
\caption[]{Percentage difference between the Improved Born Approximation
and the exact one-loop SQCD calculation for $pp
\rightarrow bh$ for minimal mixing in the
$b$ squark sector at $\sqrt{s}=7~ TeV$.}
\label{fg:minmix}
\end{center}
\end{figure}

The numerical results for $pp\rightarrow b ({\overline b})h$ at $\sqrt{s}
= 7~TeV$ were presented in (\cite{Dawson:2011pe}). The renormalization and
factorization scales were chosen to be $\mu_R=\mu_F=M_h/2$ and 
the CTEQ6m NLO parton distribution functions (\cite{Nadolsky:2008zw}) were used.
 Figs. \ref{fg:maxmix}, \ref{fg:maxmix250}
 and \ref{fg:minmix} show the percentage deviation of
 the complete one-loop
SQCD calculation from the Improved Born Approximation of Eq. 
\ref{sigibadef} for
$\tan\beta=40$ and $\tan\beta=20$ and representative values of the MSSM 
parameters.  
In both
extremes of $b$ squark mixing, the Improved Born Approximation
 approximation is within a
few percent of the complete one-loop
SQCD calculation and so is a reliable prediction for
the rate. This is true for both large and small $M_A$.
 In addition, the large $M_S$ expansion accurately 
reproduces the full SQCD one-loop result to within a few percent.
 These results are expected from the expansions of Eqs. \ref{maxans}
 and \ref{minansdef}, since
the terms which differ between the Improved Born Approximation and the
one-loop calculation are suppressed in the large $M_S$ limit.    

\begin{figure}[h]
\begin{center}
\vspace{2mm}
\includegraphics[scale=0.45]{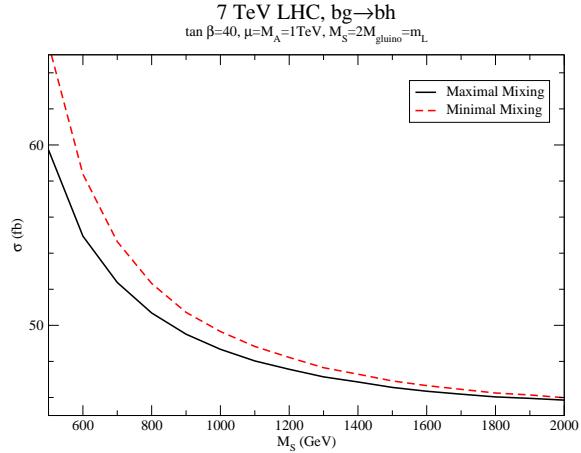} 
\caption[]{Comparison between the exact one-loop SQCD calculation for $pp
\rightarrow bh$ for minimal and maximal mixing in the
$b$ squark sector at $\sqrt{s}=7~ TeV$ and $\tan\beta=40$.
The minimal mixing curve has $m_R=\sqrt{2}M_S$ and 
${\tilde{\theta}}_b\sim 0$, while
the maximal mixing curve has $m_R=M_S$ and 
${\tilde{\theta}}_b\sim {\pi\over 4}$. }
\label{fg:compsig1}
\end{center}
\end{figure}

\begin{figure}[h]
\vspace{2mm}
\begin{center}
\includegraphics[scale=0.45]{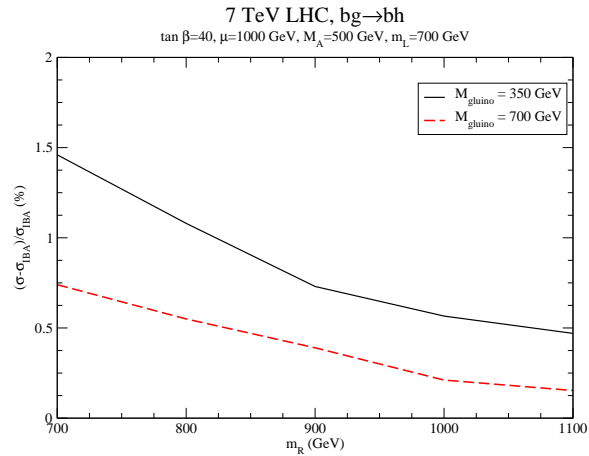} 
\caption[]{
Percentage difference between the Improved Born Approximation
and the exact one-loop SQCD calculation for $pp
\rightarrow bh$ as a function of $m_R$
 at $\sqrt{s}=7~ TeV$ and $\tan\beta=40$.
}
\label{fg:compsig2}
\end{center}
\end{figure}

Fig. \ref{fg:compsig1} compares the total SQCD rate for maximal and minimal
mixing, which bracket the allowed mixing possibilities.  For large $M_S$,
the effect of the mixing is quite small, while for $M_S\sim 800~GeV$, the 
mixing effects are at most a few $fb$.  The accuracy of the Improved Born
Approximation as a function of $m_R$ is shown in Fig. \ref{fg:compsig2}
for fixed $M_A,\mu$, and $m_L$. As $m_R$ is increased, the effects 
become very tiny.  Even for light gluino masses, the Improved
Born Approximation reproduces the exact SQCD result to within a few percent.

\begin{figure}[h]
\begin{center}
\vspace{2mm}
\includegraphics[scale=0.45]{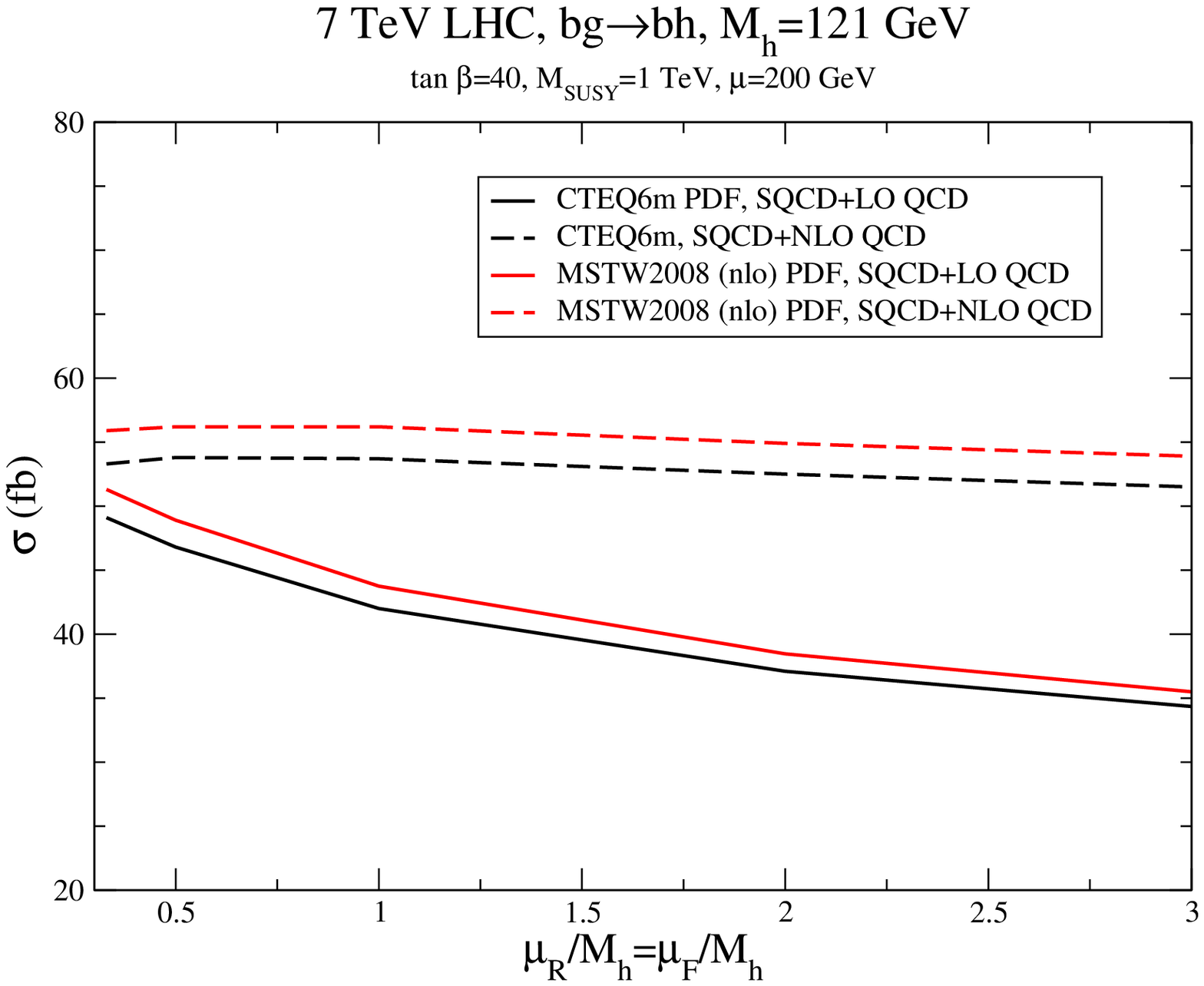} 
\caption[]{Total cross section for $pp\rightarrow b (\overline{b})h$
production including NLO QCD and SQCD corrections (dotted lines) as 
a function of renormalization/factorization scale using CTEQ6m (black) 
and MSTW2008 NLO (red) PDFs. We take $M_{\tilde g}=1~TeV$ and the
remaining MSSM parameters as in Fig. \ref{fg:maxmix}.}
\label{fg:susybh}
\end{center}
\end{figure}

\begin{figure}[h]
\begin{center}
\vspace{2mm}
\includegraphics[scale=0.45]{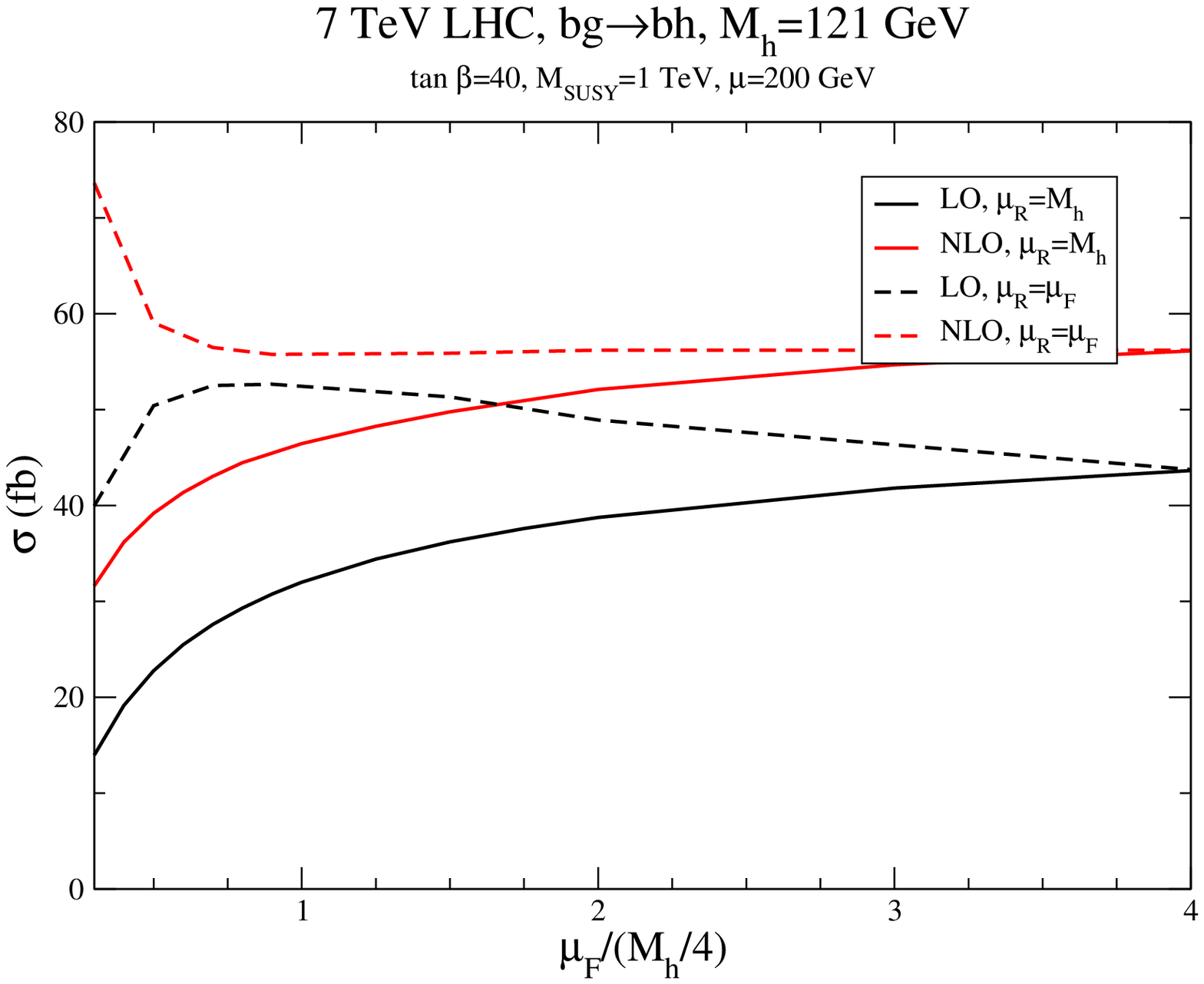} 
\caption[]{Total cross section for $pp\rightarrow b (\overline{b})h$
production including NLO QCD and SQCD corrections  as 
a function of the factorization scale using 
MSTW2008 NLO  PDFs. We take $M_{\tilde g}=1~TeV$ and the
remaining MSSM parameters as in Fig. \ref{fg:maxmix}.}
\label{fg:susybh_muf}
\end{center}
\end{figure}

In Fig. \ref{fg:susybh}, we show the scale dependence for
the total rate, including NLO QCD and SQCD corrections (dotted lines) for
a representative set of MSSM parameters at $\sqrt{s}=7~TeV$.  The
NLO scale dependence is quite small when $\mu_R=\mu_F\sim M_h$. 
 However, there is a roughly
$\sim 5\%$ difference between the predictions found 
using the CTEQ6m PDFs and
the MSTW2008 NLO PDFs\cite{Martin:2009iq}. In Fig. \ref{fg:susybh_muf},
we show the scale dependence for small $\mu_F$ (as preferred
 by \cite{Maltoni:2005wd}), and see that it is
significantly larger than in Fig. \ref{fg:susybh}. This is consistent with
the results of \cite{Harlander:2003ai,Dittmaier:2011ti}.

\section{Conclusion}
The analytical and numerical results presented in the previous sections clearly 
demonstrate that deviations from the $\Delta_b$ approximation are
suppressed by powers of $(M_{EW}/M_S)$ in the large $\tan \beta$ region.
The $\Delta_b$ approximation hence yields an accurate prediction 
in the $5$ flavor number scheme for
the cross section for squark and gluino masses at the TeV scale.

\bigskip 




\bibliography{dpf}

\end{document}